# Simple Subroutine for Inhomogeneous Deployment

Mouhamed Abdulla
Department of Electrical and Computer Engineering, Concordia University, Montréal, Québec, Canada
Email: ma14@ieee.org

*Abstract*—Spatial modeling of wireless networks via analytical means has been considered as a widely practiced mechanism for inference. As a result, some geometrical deployment models have been proposed in literature. Although practical in certain simulation instances, these models do not always produce inhomogeneous nodal geometries in an effective and simple manner for particular deployment situations. Therefore, we conceptualized a flexible approach for realizing random inhomogeneity by proposing the area-specific deployment (ASD) algorithm, which takes into account the clustering tendency of users. Overall, the developed spatial-level network tool has the distinct advantage of *automatically* producing *infinitely* many random realizations of users' geometry by simply entering three parameters to the simulator: the size of the cellular network, the number of deployment layers, and the overall quantity of nodes.

## I. INTRODUCTION

Having the capability to duplicate via models the footprint of wireless networks is important for effective network design and planning during both pre- and post-deployments. As a consequence, various inhomogeneous spatial deployment models have been suggested in literature. For instance, the principle of thinning can be applied as one possible approach, where an inhomogeneous spatial distribution is synthetically realized by deleting nodes from a uniformly deployed pattern [1]. Another technique enables heterogeneity through different adaptation of edge or center-focused deployments by adjusting the spatial models through a tunable variable [2], [3]. Additionally, Gaussian geometry is yet another interesting heterogeneous model because both the geographical spread and the intensity of terminals are changeable [3], [4].

Although applicable for preliminary analysis, these spatial deployment models and the like will not necessarily generate reliable mapping of sporadically positioned inhomogeneous networks. Thus, it becomes interesting to explore new practices for inhomogeneous deployment. In this paper, we intend to develop the ASD algorithm, which is an adaptable deployment mechanism that will require limited *a priori* input parameters from designers. Hence, the simulator can easily be applied and configured for an array of network planning projects.

The rest of this paper is organized as follows. In Section II, we will explain the logic behind the ASD algorithm for the purpose of emulating spatial inhomogeneity. Then, in Section III, we will utilize this proposed algorithm to develop a heterogeneous mechanism for *automatic* random deployment that can be initialized with only three simple geometrical parameters: the size of the network, the number of deployment layers, and the overall number of nodes. Finally, Section IV will conclude the paper.

## II. ASD DEPLOYMENT STRATEGY FOR HETEROGENEITY

Conceiving an algorithm for inhomogeneity that can be configured easily may be realized by contriving a framework that adheres to the notion of *divide and conquer*. That is, deployment can be tackled by breaking-down this task into smaller algorithmically solvable parts, and then synthesizing the results. In particular, this is done by proposing a superposition-based algorithm which we refer to as ASD. As shown in the representative example of Fig. 1, the ASD approach is systematic. Essentially, for a particular project site, the designer will identify likely clusters in a way that the overall network scale and surface is split among these sub-regions. As a matter of fact, each of the non-overlapping clusters is uniquely specified by its: support domain $D_i \subseteq \mathbb{R}^2$; surface area $A_i \in \mathbb{R}_+^*$; quantity of randomly positioned nodes $n_i \in \mathbb{N}^*$; and corresponding number density $\rho_i \in \mathbb{R}_+^*$. Moreover, the areal size and nodal scale of the original lattice are respectively given by: $A_N = \sum_{i=1}^{n_{\text{sec-total}}} A_i$ and $n_S = \sum_{i=1}^{n_{\text{sec-total}}} n_i$, such that $n_{\text{sec-total}} \in \mathbb{N}^*$ is the overall amount of sectors.

Once the planning of the network is set, we then focus on the sub-regions in a standalone way so as to stochastically generate the desired number of positions. Random uniform deployment is assumed over the specified sectors as expressed by the particular density function of the sub-regions: $f_{XY}(x,y) = 1/A_i \cdot \mathbf{1}_{D_i \subseteq \mathbb{R}^2}(x,y) \quad i = 1, 2, \cdots, n_{\text{sec-total}}$. Following the analysis step, the various subnetworks are reassembled in a puzzle-like format; thus resulting spatial heterogeneity.

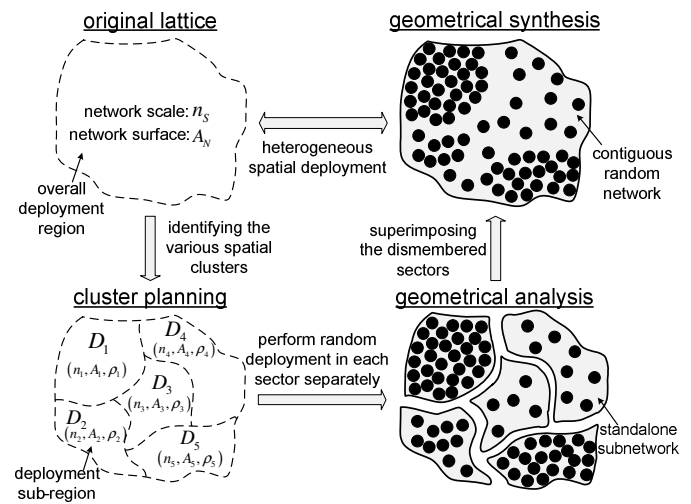

Figure 1. Characterizing ASD by a descriptive network model.

## III. AUTOMATIC ALGORITHM FOR RANDOM DEPLOYMENT

In previous work, we developed a simulator tool for inhomogeneity based on controlled network planning [5], [6]. While the approach is adequate, in some cases, we want to provide greater emulation leverage to the designer by conceptualizing another algorithm that can achieve heterogeneity with very limited planning information supplied to the network subroutine. Thus, the aim here is to construct an inhomogeneous random network in an automatic manner.

As explained in Section II, inhomogeneity is possible by having different areal number densities in each of the deployment sub-regions. Clearly, the density for the sectors is obtained by: $\forall A_i \in \mathbb{R}_+^* : \exists n_i \in \mathbb{N}^* : \rho_i \triangleq n_i/A_i \quad i=1,2,\cdots,n_{\text{sec-total}}$. Therefore, it is possible to have unique densities by varying $A_i$, and keeping $n_i$ fixed.

Next, in order to obtain different sub-regions, we will consider $n_L \in \mathbb{N}^*$ onion-like layers; therefore, for this layout $n_{\text{sec-total}} = n_L$. In fact, for the purpose of automatic inhomogeneity, the number of layers will randomly be chosen from a predefined integer range: $\Pr\{X=x\} = \mathcal{U}_D(n_1, n_2) \quad (n_1, n_2) \in \mathbb{Z}^2 : n_1 \leq n_2$. Indeed, we aim to randomly generate $n_L$ from a range delimited by $n_{L-\max}$, which essentially refers to the maximum arbitrary number of deployment layers possible for achieving spatial inhomogeneity. This value will actually be preset by the network designer at the start of the automatic emulation process. Thus, the number of layers at a simulation instance will be given by: $n_L \sim \mathcal{U}_D(2, n_{L-\max}) \quad n_{L-\max} \in \mathbb{N}^* : n_{L-\max} > 1$. Notice that the sampling range begins at $n_1 = 2$ because from the ASD principle, we at least need 2-layers for attaining inhomogeneity.

The next step requires us to equally split the number of nodes among these sub-regions. By design, the overall amount of nodes $n_S \in \mathbb{N}^*$ planned for deployment is supplied by the network architect. Since $n_S$ and $n_L$ need not necessarily be multiples of each other, then the number of nodes per layer must be arranged in a careful way. In particular, the amount of random nodes deployed in the innermost layer of an automatically emulated inhomogeneous network is designated by $n_{in} \in \mathbb{N}^*$. As for the outer layers, each of these sub-regions will contain $n_{out} \in \mathbb{N}^*$ nodes computed by: $\forall n_S \in \mathbb{N}^* : \exists n_L \in \mathbb{N}^* : n_{out} \triangleq \lfloor n_S/n_L \rfloor$. Knowing the volume of nodes in the outer layers, then it should be evident that the rest of the overall nodal quantity will constitute the amount of terminals in the innermost sub-region of the cell. Therefore, this measure can be calculated as follows:

$$n_{in} \triangleq n_S - (n_L - 1) \cdot n_{out} = n_S - (n_L - 1) \cdot \lfloor n_S/n_L \rfloor \quad (1)$$

So far, we have determined the number of layers and the amount of nodes in each sector. At present, we want to vary the areal size of each sub-region. This task can be done by randomly deciding on the geometrical position of the layers.

That is, we want the width or thickness $\Delta_i \in \mathbb{R}_+^*$ of the various deployment layers to be different. In fact, this value corresponds to: $\Delta_i = r_i - r_{i-1} \quad i = 2, 3, \cdots, n_L$, such that $\Delta_1 = r_1$ is the radius measured from the origin of the Cartesian coordinate system to the first layer, and $r_i \in \mathbb{R}_+^*$ is the particular radius for all the other deployment layers. In this situation, the procedure to generate diverse widths of the deployment sub-regions can be realized by randomly producing radial values for the layers; this can be accomplished by: $r_i \sim \mathcal{U}_R(0, L) \quad i = 1, 2, \cdots, n_L - 1$.

Following the generation of these radial distances, it then becomes necessary to sort them in ascending order, i.e.: $\vec{r}_{sorted} = \text{sort}(\vec{r} \in \mathbb{R}_{+,*}^{n_L-1}) = \text{sort}([r_i]_{i=1,2,\cdots,n_L-1})$. There are many techniques available for implementing the sorting operator; quicksort has been established as one of the fastest algorithms for ordering an array of numbers.

Next, we will stochastically deploy in each of the formed random sized sub-regions the corresponding amount of nodes. Then, we superimpose these multi-density sectors together and look at the network as a holistic entity, which results into an inhomogeneous spatial structure. To be precise, the geometrical randomness is achieved due to the amalgamated arbitrary nature of: the number of deployment layers $n_L$, the size of the layers $\Delta_i$, and the position of nodes within each layer $\{\hat{x}_i, \hat{y}_i\}$. For the sake of completeness, these attributes are graphically depicted in the geometrical model of Fig. 2 used for automatically producing a random network footprint.

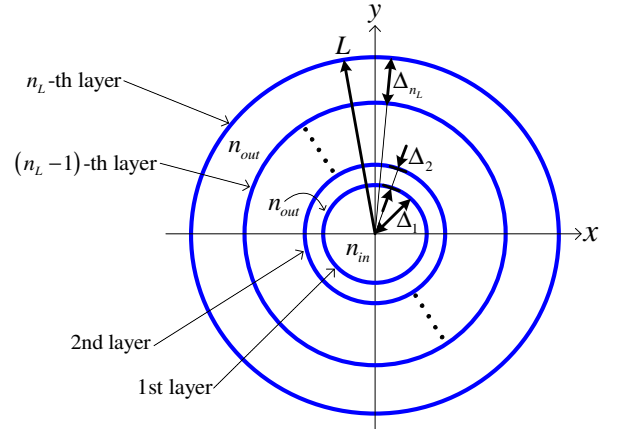

Figure 2. Geometrical details for automatic random deployment.

Overall, the culmination of the above explanations and analysis enables us to derive the automatic inhomogeneous algorithm of Fig. 3. It can vividly be observed that a designer will only require entering three essential inputs: the size of the cellular network $L$, the maximum number of deployment layers $n_{L-\max}$, and the quantity of nodes to be deployed $n_S$.

Meanwhile, the time performance of the above algorithm is evaluated in (2). In this cost analysis, we considered $n_L = n_{L-\max}$ so as to reflect the worst computational scenario.

$$O(T_{\cos t}(L, n_{L-\max}, n_S)) \sim O(n_S + n_{L-\max} \cdot \log_2 n_{L-\max}) \quad (2)$$

## Algorithm 1 - Emulating an Inhomogeneous Random Network - Automatic Deployment

1: Require: $n_S \in \mathbb{N}^*$ $\quad L \in \mathbb{R}_+^*$ $\quad \{n_{L-\max} \in \mathbb{N}^* | n_{L-\max} > 1\}$
2: Generate a RV: $\hat{u}_0 \sim \mathcal{U}(0,1)$
3: Compute: $\hat{v}_0 := \{3/2 + \hat{u}_0(n_{L-\max} - 1)\} \sim \mathcal{U}_V(3/2, n_{L-\max} + 1/2)$
4: **for** $i = 2, 3, \cdots, n_{L-\max}$ **do**
5:   **if** $\{|\hat{v}_0 - i| \leq 1/2\}$ **then**
6:     $n_L := i \sim \mathcal{U}_D(2, n_{L-\max})$ **break for**
7:   **end if**
8: **end for**
9: Compute: $n_{in} := n_S - (n_L - 1) \cdot \lfloor n_S/n_L \rfloor \quad n_{out} := \lfloor n_S/n_L \rfloor$
10: **for** $j = 1, 2, \cdots, n_L - 1$ **do**
11:   Generate a RV: $\hat{u}_1 \sim \mathcal{U}(0,1)$
12:   Compute: $\hat{r}_j := \hat{u}_1 L \sim \mathcal{U}_R(0, L)$
13: **end for**
14: Sort in ascending order: $\{\hat{r}_j\} := \text{sort}(\hat{r}_j) \quad j = 1, 2, \cdots, n_L - 1$
15: Initialize: $t = 0$
16: **for** $j = 1, 2, \cdots, n_L$ **do**
17:   **if** $\{j \neq 1\}$ **then**
18:     $L_1 := \hat{r}_{j-1} \quad n_0 := n_{out}$
19:   **else**
20:     $L_1 := 0 \quad n_0 := n_{in}$
21:   **end if**
22:   **if** $\{j \neq n_L\}$ **then** $L_2 := \hat{r}_j$ **else** $L_2 := L$ **end if**
23:   **for** $m = 1, 2, \cdots, n_0$ **do**
24:     Generate two i.i.d. RVs: $\{\hat{u}_2, \hat{u}_3\} \sim \mathcal{U}(0,1)$
25:     Compute: $\hat{r}_m := \sqrt{L_1^2 + \hat{u}_2(L_2^2 - L_1^2)} \sim f_R(r)$
26:     Compute: $\hat{\theta}_m := 2\pi \hat{u}_3 \sim f_\theta(\theta)$
27:     Compute: $\{\hat{x}_{t+m}; \hat{y}_{t+m}\} = \{\hat{r}_m \cos(\hat{\theta}_m); \hat{r}_m \sin(\hat{\theta}_m)\} \sim f_{XY}(x,y)$
28:   **end for**
29:   $t := t + n_0$
30: **end for**
31: Return: $\{\hat{x}_i, \hat{y}_i\} : i = 1, 2, \cdots, n_S$

Figure 3. Pseudocode for automatic heterogeneous spatial deployment.

In Fig. 4, we show the Monte Carlo simulations of four random instances of a small-scale heterogeneous deployment based on the following parameters: $(L, n_{L-\max}, n_S) = (1, 5, 100)$.

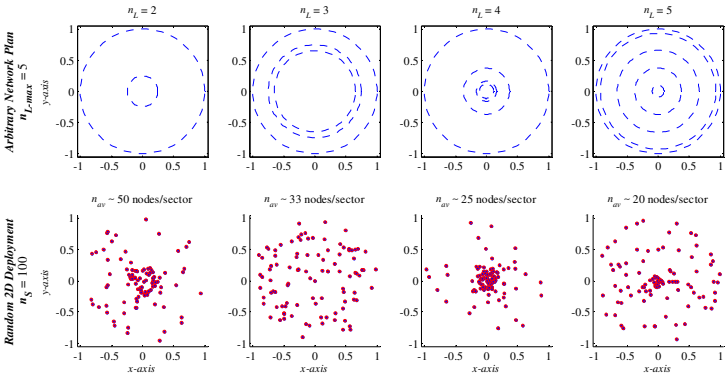

Figure 4. Random instances of small-scale heterogeneous deployment.

Similarly, in Fig. 5, we demonstrate another set of simulation examples for a medium-scale network with: $(L, n_{L-\max}, n_S) = (1, 10, 1000)$. Again, each run of the simulation automatically produces a unique inhomogeneous realization.

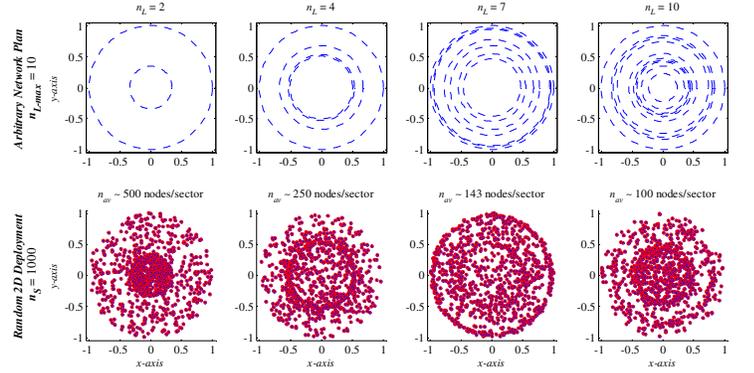

Figure 5. Random instances of medium-scale heterogeneous deployment.

## IV. CONCLUSION

In this paper, we remarked that typical deployment methods, though practical to some degree, have their limitations. Thus, an inhomogeneous deployment algorithm based on the superposition principle of targeted spatial distribution was proposed. This conceptualized approach, which we refer to as ASD, is more manageable because it breaks-down the complicated task of finding the wholesome distribution of users' spatial pattern in a terrain to that of smaller sub-regions. Then, the principle of superposition is applied to merge the spatial clusters together, and hence establish the entire random distribution of the network so as to investigate various integrity measures.

Overall, the key advantage of the developed inhomogeneous algorithm is that it can *automatically* construct a *unique* random deployment pattern while necessitating only 3 basic input parameters, namely: the size of the cellular network, the number of deployment layers, and the overall quantity of nodes.